\documentclass[11pt]{article}
 \usepackage[letterpaper,hmargin=1in,vmargin=1.25in]{geometry}

\begin{document}

\title{Using Repeating Decimals As An Alternative To Prime Numbers In Encryption}


\author{Givon Zirkind}




\date{June 22, 2010}


\begin{abstract}
This article is meant to provide an additional point of view, applying known knowledge, to supply keys that have a series of non-repeating digits, in a manner that is not usually thought of.   Traditionally, prime numbers are used in encryption as keys that have non-repeating sequences.   Non-repetition of digits in a key is very sought after in encryption.  Uniqueness in a digit sequence defeats decryption by method.  In searching for methods of non-decryptable encryption as well as ways to provide unique sequences, other than using prime numbers, the idea of using repeating decimals came to me. Applied correctly, a repeating decimal series of sufficient length will stand in as well for a prime number.  This is so, because only numbers prime to each other will produce repeating decimals and; within the repeating sequence there is uniqueness of digit sequence.
\end{abstract}

\maketitle



\section{Introduction}

\indent\indent Prime numbers are often used and sought after, in encryption.  One of the several  reasons for this is, that prime numbers can be used as keys that have a sequence of non-repeating digits.  [2]  (This is a fundamental concept of encryption and a pre-requisite for understanding the application of the number theory disccused in this paper.  This fundamental concept of encryption will not be explained in this paper.  The author refers the reader to [2] for an in depth explanation of this concept.)  One of the many other ways of producing a sequence of non-repeating digits, besides using prime numbers, is to use the sequences of repeating decimals.  Like prime numbers, repeating decimals, within their sequence, do not have repeating (sub)sequences.  \newline

Ex.  $\frac{1}{7}$  = 0.142857142857…  within the repeating sequence “142857”, there is uniqueness.  Similar to a prime number.  [3] \newline

Furthermore, for any given message, of any size, we have a formula, Fermat’s Little Theorem, to select a denominator that will produce a repeating sequence with a specific number of digits.  [10$^($$^p$$^-$$^1$$^)$ $\equiv$ 1 mod p]   [3] [4]   All one has to do, is select a denominator, that will produce a repeating sequence with at least as many digits as there are characters in the message to encrypt.  Then, an appropriate key with no repetition or pattern in its sequence of digits has been generated. \newline

We also know the rules to producing a repeating decimal.  Such as, the denominator can not have a power of 2 and/or 5.  [2]    Also, the numerator and denominator must be prime to each other.  They must have no common factors.  They need not necessarily be prime.  \newline

Ex. $\frac{4}{9}$=0.444…  While neither 4 nor 9 is prime, they are prime to each other.  Hence, they can never fully divide each other.  [1] [5]  \newline

So, although we wish we could generate prime numbers and can not; we can generate repeating decimals and choose their size.  All we have to do, is choose an appropriate size and select the appropriate factors.  This makes the use of repeating sequences an attractive option for key generation.    \newline


\section{The Infinity of  Repeating Decimals}

\subsection{The Number of Prime Numbers Is Infinite \newline The Number of Repeating Decimals Is Infinite}
\indent \indent The number of prime numbers is infinite.  [6]   \  Every prime number will generate several repeating decimals.   So, the number of repeating decimals generated by prime numbers is infinite.   \newline

\subsection{The Number of Repeating Decimals Is Greater Than The Number of Prime Numbers}
\indent \indent In  fact, as there are many repeating decimals generated by each prime number, there is a greater than a one-to-one correspondence between primes and repeating decimals generated by prime numbers.  Hence, the number of repeating decimals is greater than the number of prime numbers. \newline


\subsection{The Number of Co-Primes Numbers Are Infinite \& \newline Greater Than The Number of Prime Numbers}
\indent \indent There are more co-primes than primes.  Because there are more odd numbers than prime numbers.  For while every prime number is odd, not every odd number is prime.  [5]  \ And, every pair of odd numbers that do not share another odd number as a factor are co-prime.   This includes all prime numbers (which are co-prime with each other) \textbf{\emph{plus}} all odd multiples of prime numbers that do not share the same multipliers \textbf{\emph{and}} any pairs of combination of primes and multipliers that do not share the same primes and multipliers.  Therefore, the number of co-primes is greater than the number of primes.  \newline

\indent Also, even and odd numbers are co-prime; while prime numbers can be only odd numbers.  Therefore, there are more co-prime numbers than prime numbers.   And, since prime numbers are infinite and; there are more co-prime numbers than prime numbers; therefore co-prime numbers are also infiinite.\newline

\indent In addition, since the number of co-primes are greater than the number of primes and; since every pair of co-primes produces a repeating decimal, therefore, there are more repeating decimals than prime numbers (and the number of repeating decimals is infinite as deduced above).    \newline

\section{Analysis and Examples of Co-Primes}

\subsection{How Odd Numbers Share Prime Factors}
\indent \indent It is clear how many pairs of odd numbers share common factors.  One way of demonstrating this is to analyze the series of odd numbers and their corresponding ordinal numbers.\newline

\subsection{Analysis of the Series of Odd Numbers}

\noindent
\begin{tabular}{|c|| *{19}{c|}}
\hline
\multicolumn{20}{|c|}{Series of Odd Numbers}\\
\hline
Position&1& 2 & 3 &4 &5 &6 &7 &8 &9 &10 &11 &12 &13 &14 &15 &16 &17 &18  &19\\
\hline
Odd Number&3& 5& 7 &9 &11 &13& 15& 17& 19& 21& 23& 25& 27& 29& 31& 33& 35& 37 &39\\
\hline
\end{tabular}
\newline
\newline

Every odd number will share a common factor with every other odd number in the series of odd numbers, whose ordinal number is a multiple of itself plus one.  \newline

Ex.  The number 3 is the first odd number.  It will share a factor with the 4th odd number as $4 =(1 *  3) + 1$.  The 4th odd number is 9.  The 7th odd number, $7 = (2*3) + 1$, is 15.  This too has a common factor with 3.  \newline

Ex.  The number 39 is the 19$^{th}$ ordinal odd number.  19 = (6$^{th}$*3) + 1 and; (3$^{rd}$*6) +1.  The number 39 has the 1st and 6th ordinal odd numbers as factors or:  39 = 13 * 3. \newline

\subsection{The Number of Co-Primes for a Given Odd Number}
For the 1$^{st}$ odd number, 3, within every period of 3 numbers, will have 2 co-prime numbers.   Three will be co-prime with $\frac{2}{3}$ of the series of odd numbers.  The 2$^{nd}$ odd number, 5, within every period of 5 numbers, will have 4 co-prime numbers.  Five will be co-prime with $\frac{4}{5}$ of the series of odd numbers.  And, so forth.  Odd numbers that are multiples of a given odd number, will reduce to a smaller fraction.  Thus reducing the number of co-primes for a given odd number.  The end result will be that the number of co-primes for any odd number 'n' is equal to [$\frac{(n-1)}{n}$ * O].   \ Where 'O' equals the number of odd numbers. \newline

\subsection{The Total Number of Co-Primes for All Odd Numbers}
\indent \indent As the number 'n' of odd numbers increases, the number of pairs of numbers co-prime (n$_{cp}$) with an odd number approaches the number of odd numbers.  \\
\indent Eq.  $\frac{lim}{n\rightarrow\bowtie}$$\frac{(n-1)}{n}$ $\rightarrow$ $\frac{n}{n}$  $\rightarrow$ 1.  \\
\indent Eq.  $\frac{lim}{n\rightarrow\bowtie}$[$\frac{(n-1)}{n}$] = 1= O.   \\
\indent The sum of pairs of odd numbers that are co-primes approaches the number of odd numbers, an odd number of times.  Or; the number of odd numbers (O) squared (O$^2$).  \\
\indent Eq. $\frac{lim}{n_{cp}\rightarrow\bowtie}$ = O$^2$. \\
\indent While the number of primes (n$_p$), is less than the number odd numbers.  [n$_p$ $<$ O]. \\
\indent This also proves that  there are more co-primes than primes.  \\
\indent Eq.   [n$_p$ $<$ O] \\
\indent Eq.   [n$_{cp}$ $=$ O$^2$] \\ 
\indent Eq.   [O $<$ O$^2$] \\
\indent Eq.   [n$_p$ $<$ n$_{cp}$] \\
\newline

This calculation does not include the number of co-primes that the odd numbers will have with even numbers.  \newline

Ex.  The number 33 besides being co-prime with all odd numbers not multiples of 3 and 11; will be co-prime with every even number. \newline

\subsection{An Arithmetic Explanation of Common Factors For Odd Numbers\newline Identifying Unique Co-Primes}
In arithmetic terms:\\
\newline
Every odd number is equal to (2n+1).\\
f = The 1$^{st}$ position a particular odd number 'o' occurs within the series of odd numbers.\\
o = an odd number\\
m = a multiplier of an odd number\\
Then, for every multiple of an odd number 'o' that will be odd,  n = f + (o*m). \newline

\noindent Not every multiple of an odd number is odd.\\
\newline
Ex. 15, 
30, 45, 60, 75 are all multiples of the odd number 15.\\
But, only 15, 45 and 75 are odd.\\
Because any odd number times an even number is even.  \\
Only an odd number times an odd number is odd.\newline

\noindent Using the formula to calculate the odd multiples of 15:\newline\newline
f$_{15}$ = 7\\
o = 15\\
For m=\{1, 2, 3, 4\}\\

\noindent
\begin{tabular}{|c|c|c|c|c|c|c|}
\hline
Odd&First Position&Multiplier (m)&n=f+(o*m)&2n+1&$\frac{O}{N}$=$\frac{Reduced}{Fraction}$&$\frac{N}{O}$\\
Number&(Occurence)&&&&&\\
\hline
15 & 7 & 1 & 22=7+(15*1) & 45&$\frac{15}{45}$=$\frac{1}{3}$&3\\
\hline
15 & 7 & 2  & 37=7+(15*2) & 75&$\frac{15}{75}$=$\frac{1}{5}$&5\\
\hline
15 & 7 & 3  & 52=7+(15*3) & 105&$\frac{15}{105}$=$\frac{1}{7}$&7\\
\hline
15 & 7 & 4  & 67=7+(15*4) & 135&$\frac{15}{135}$=$\frac{1}{9}$&9\\
\hline
\end{tabular}
\\
\newline
\newline
\indent The series of odd numbers that are the multiples of an odd number, is reducible itself, to the series of odd numbers.  This is so, because only an odd number times an odd number will produce an odd number.  An even number times an odd number will be even.  Therefore, to have multiples of an odd number that are odd, one must multiply by odd numbers.  Thus, the series of the inverse of odd numbers emerges as the cyclic pattern of common factors to odd numbers now becomes apparent.  \newline

(The similarity with the Sieve of Eratosthenes [5]  is uncanny.  Although, that was not my original intention.  The similarity merely presented itself.)

\subsection{The Cyclic Pattern of Common Factors Reduces the Number of Co-Primes}

\indent \indent As demonstrated in the table above, odd numbers that are multiples of other odd numbers, reduce to the series of inverse odd numbers.  This is true for every odd number.  Thus many co-primes will produce the same repeating decimal.  This cyclic pattern of common factors to odd numbers does reduce the number of co-primes and possible repeating decimals.  Still, an infinity of co-primes will produce an infinity of repeating decimals.\newline

\subsection{A Determination Of More Co-Primes Than Primes In a Bounded Series of Odd Numbers}

\indent \indent It is obvious that for any number of integers where there are at least 2 primes greater than half the number of integers in the series, there are more co-primes than primes and; more repeating decimals than primes. \newline

Ex.  From 1 to 100, there is 51 and 67.  Each will produce a repeating decimal for numerator. That is a total of 118 different sequences.  And, there are more primes within the series from which to generate repeating sequences.  Yet, obviously, the number of primes between 1 and 100 must be less than 100.  \newline

\subsection{Non-Prime Numbers May Be Co-Prime}

\indent \indent In addition, when searching for co-primes, one must also reckon the repeating sequences from numbers that are prime to each other, but not necessarily prime.  Again, the total number of decimal repeating sequences will be larger than the number of primes within the given series of numbers. \newline

Ex.  --  Primes, Co-Primes \& Repeating Decimals Less Than 10
\begin{enumerate}
\item The number of primes less than 10 is just 4 \{1, 3, 5, 7\}.  
\item Ignore any number that is a power of 2 or 5 in the denominator  \{2, 4, 5, 8\}.  
\item The number of decimal repeating sequences using the digits from 1 -- 10, is 14, which, is greater than 4, the number of primes from 1 -- 10.  
\item The number of decimal repeating sequences from the prime numbers \{7\} that are greater than half of the number of elements in the series is 6  \{$\frac{1}{7}$, $\frac{2}{7}$, $\frac{3}{7}$, $\frac{4}{7}$, $\frac{5}{7}$, $\frac{6}{7}$\}.  
\item Note, the fractions \{$\frac{2}{7}$, $\frac{4}{7}$, $\frac{6}{7}$\} have even numbers as numerators and yet, produce repeating decimals.  Because these even numbers are co-prime with the denomiators.  

\item As demonstrated by the series of fractions produced by the prime number 7 \{$\frac{1}{7}$, $\frac{2}{7}$, $\frac{3}{7}$, $\frac{4}{7}$, $\frac{5}{7}$, $\frac{6}{7}$\}; If the prime number is greater than half the number of numbers in the range:   
	\begin{enumerate}
\item Then, the number of numerators  is also greater than half the number of numbers in the range.
\item Then, the number of numerators are greater than half the number of prime numbers in the range.
	\end{enumerate}
\item The number of repeating decimals from numbers prime to each other with a non-prime denomiator is 6  \{$\frac{1}{9}$, $\frac{2}{9}$, $\frac{4}{9}$, $\frac{5}{9}$, $\frac{7}{9}$, $\frac{8}{9}$\}.
\indent	This is also greater than the number of primes in the series.  
\item The total number of repeating sequences is 14  \{$\frac{1}{3}$, $\frac{2}{3}$, $\frac{1}{7}$, $\frac{2}{7}$, $\frac{3}{7}$, $\frac{4}{7}$, $\frac{5}{7}$, $\frac{6}{7}$, $\frac{1}{9}$, $\frac{2}{9}$, $\frac{4}{9}$, $\frac{5}{9}$, $\frac{7}{9}$, $\frac{8}{9}$\} 
\indent This more than 3 times the number of primes in the series. 
\item The repetitive fractions $\frac{3}{9}$ and $\frac{6}{9}$ are redundant to $\frac{1}{3}$ and $\frac{2}{3}$ which have already been counted.  These fractions are not counted twice.\newline
\end{enumerate}

\subsection{Multi-Factored Numbers That Are Co-Prime}
\indent \indent In addition, the entire discussion above, is only of a pair of single factor numbers that are co-prime.  One can also use multiplicands of odd numbers that are co-prime. Thus, another infinity of co-primes can be generated.\newline

\noindent Ex.  \newline
$[3 $/$ (17 *23)]$  or \\
$[( 4 $/$ 27)]$ or  \\
$[ ( 3 * 11) $/$ (5 * 13)]$ or \\
$[(3 * 5 * 7) $/$ (11 * 13 * 17)]$ or \\
$[64 $/$ (11 * 13 * 17)]$\\

\section{Conclusion}
\indent \indent Not all the repeating decimals will be of use for encryption.  In fact, many will be useless for encryption.  However, it becomes apparent that there are enough repeating decimal sequences to provide quality encryption.   It is also apparent that there are more repeating decimals that will provide non-repetitive digit sequences for keys, than there are prime numbers that will provide non-repetitive digit sequences for keys.  \newline

Most messages are  500 characters or less.  (This is a statistic from standard radio-telegraphy and studies of typing.)  To generate a unique key, for encryption, with enough digits, that do not repeat, for a message of 500 characters, only requires a prime number with as many digits as the message.  A repeating decimal sequence of 500 characters or less, can be easily generated.  It too will not have repeating sequences within it.  It will contain a unique series of non-repeating digits.  It will be just like a prime number in that regard. \newline

Also, using the unique sequences of repeating decimals adds another possibility to check for, when deciphering.  (Presumably, decryption methods worth their salt, automatically check for prime numbers.)  For individual and low grade traffic, using a repeating sequence as a key, would be an acceptable, secure method of encryption. \newline

\section{Further Research}

\subsection{}
To form an equation for the total number of co-primes within a given range of numbers.  (This is a current work in progress.)

\subsection{}
The number of odd numbers and prime numbers have the transfinite number of $\aleph$$_0$.  Does the number of co-primes also have the transfinite number of $\aleph$$_0$?  More importantly, is the number of repeating decimals from co-primes equal to $\aleph$$_0$?  Or, is the transfinite number of the repeating decimals from co-primes greater than $\aleph$$_0$ since there is a one-to-many relationship of repeating decimals to a pair or group of co-primes?

\section{Bibliography}

[1]  Euclid, Book VII, Proposition 1 \& 29  \newline

\noindent [2]  Kahn, David; The Codebreakers: The Story of Secret Writing; Scribner, New York, 1996; ISBN 0684831309  \newline

\noindent [3]  Wikidepia.Org; Repeating Decimal  \newline

\noindent [4]  Carl Friedrich Gauss, On The Congruence of Numbers, Translated by Ralph G. Archibald, From Disquisitiones Artimeticae; Reprinted in Treasury of Mathematics, Edited by Henrietta O. Midonick, Philosophical Library Inc., 1965  \newline

\noindent [5]  Nichomachus of Gerasa, Introduction to Arithmetic, Translated by Martin Luther D'Ooge; Reprinted in Treasury of Mathematics, Edited by Henrietta O. Midonick, Philosophical Library Inc., 1965  \newline

\noindent [6] Euclid, Book VII


\bibliographystyle{amsplain}
\bibliography{gz-jams-l}


\end{document}